\documentclass[useAMS,usenatbib]{mnras}
\usepackage{epsfig,amsmath,amssymb,amsfonts,mathrsfs,latexsym,graphicx}

\usepackage{hyperref}
\usepackage{natbib}
\usepackage{color}
\usepackage{algorithm}
\usepackage{ textcomp } 

\usepackage{fixltx2e}

\usepackage{float}
\usepackage[caption = false]{subfig}
\usepackage{multirow}
\include{journals}

\graphicspath{{images/}}

\newcommand{\kms}{km~s$^{-1}$}

\newcommand{\SiII}{Si~{\sc ii}}

\newcommand{\Deltam}{$\Delta m_{15}(B)$}

\def\gsim{\mathrel{\rlap{\lower 4pt \hbox{\hskip 1pt $\sim$}}\raise 1pt \hbox {$>$}}}
\def\lsim{\mathrel{\rlap{\lower 4pt \hbox{\hskip 1pt $\sim$}}\raise 1pt \hbox {$<$}}}

\title[Breaking the color-reddening degeneracy in type Ia supernovae]{Breaking the color-reddening degeneracy in type Ia supernovae}

\author[M. Sasdelli et al.]
{\parbox{\textwidth}{\vspace{-.5cm}Michele~Sasdelli$^{1,2}$\thanks{E-mail: 
sasdelli@m.sasdelli@ljmu.ac.uk}, 
E.~E.~O.~Ishida$^{3,2}$,
 W.~Hillebrandt$^{2}$,
C.~Ashall$^{1}$,
P.~A.~Mazzali$^{1,2}$,
S.~Prentice$^{1}$
}\vspace{0.6cm}\\
\parbox{\textwidth}{ 
$^{1}$Astrophysics Research Institute, Liverpool John Moores University,
Liverpool L3 5RF, UK \\
$^{2}$Max-Planck-Institut f\"ur Astrophysik, Karl-Schwarzschild-Str. 1, 85741 Garching bei M\"unchen, Germany\\
$^{3}$Clermont Universit\'e, Universit\'e Blaise Pascal, CNRS/IN2P3, Laboratoire de Physique Corpusculaire, BP 10448, F-63000 \\ Clermont-Ferrand, France\\
}}

\begin{document}

\date{Accepted ... Received ...; in original form ...}

\pagerange{\pageref{firstpage}--\pageref{lastpage}} \pubyear{2016}

\maketitle
\label{firstpage}

\begin{abstract}

A new method to study the intrinsic color and luminosity of type Ia supernovae
(SNe\,Ia) is presented.  A metric space built using principal component
analysis (PCA) on spectral series SNe\,Ia between $-12.5$ and $+17.5$ days from
$B$ maximum is used as a set of predictors. This metric space is built to be insensitive to
reddening.  Hence, it does not predict the part of color excess due to dust-extinction. At the
same time, the rich variability of SN\,Ia spectra is a good predictor of a
large fraction of the intrinsic color variability.  Such metric
space is a good predictor of the epoch when the maximum in the $B-V$ color curve is reached.
Multivariate Partial Least Square (PLS) regression predicts the intrinsic $B$ band light-curve and the intrinsic
 $B-V$ color curve up to a month after maximum.  This allows to
study the relation between the light curves of SNe\,Ia and their spectra.  The
total-to-selective extinction ratio R$_V$ in the host-galaxy of SNe\,Ia is
found, on average, to be consistent with typical Milky-Way values.
 This analysis shows the importance of
collecting spectra to study SNe\,Ia, even with large sample publicly available.
  Future automated surveys as
LSST will provide a large number of light curves.  The analysis shows that
observing accompaning spectra for a significative number of SNe will be
important even in the case of ``normal'' SNe\,Ia.

\end{abstract}


\section{Introduction}
\label{sec:intro}

The use of type Ia supernovae (SNIa) as standardizable candles  provided the
first evidence of an accelerated expansion of the Universe in late 20$^{th}$
century \citep{1998AJ....116.1009R,1999ApJ...517..565P}. Since these first results were presented the
standardization procedure improved considerably, allowing for tighter
constraints over cosmological parameters. Consequently, we have reached a point
where the intrinsic characteristics of the SNe can not be neglected if we wish
to improve the calibration and take into account systematic effects. The
colour-dust connection is one such bottleneck. 

The typical total to selective extinction estimate from Milky-Way measurements,
 $R_V=3.1$, is significantly different from the one obtained from SNIa Hubble
residuals, $1.7<R_V<2.5$.
This discrepancy has been attributed to a poor understanding of the intrinsic
color variability or to unusual extragalactic or interstellar dust. 
Disentangling extrinsic and intrinsic variations in SNIa samples is a difficult
and essential problem, which might impact calibration for all subsequent
cosmological parameters estimations\footnote{Here we consider
\textit{intrinsic} any effect present in the circumstellar medium of the
progenitor which cannot be observationally identified as a separate effect.}.

Chotard et al., 2011 showed that corrections of the intrinsic color derived for
individual analysis of equivalent widths may lead to an $R_V$ estimated from
SNIa which agrees with standard Milky-Way values. In this paper, we improved
upon the techniques presented in \cite{2015MNRAS.447.1247S} 
and use the correlations between spectral series and light curves to study the
behaviour of the $B-V$ color curve and the host-galaxy total-to-selective
extinction ratio $R_V$ of SNe\,Ia.  Our method uses a data set of spectra
series and light curves in order to estimate a color law for SNe\,Ia. Once this
is determined, the identification of how much of the color is due to dust and how
much of it is due to intrinsic properties is straightforward for an
object with spectroscopic observations.

The core argument presented in  \cite{2015MNRAS.447.1247S} relies on finding correlations between
spectral series and light curve measurements on the same set of SNe\,Ia. 
The radiation
transport physics that forms the spectra is a complex phenomenon.
A lot of
information on the physical structure of the ejecta is encoded 
in its spectral features and their time evolution.
 However, using this information in the spectra to measure the amount of extinction and the type of dust along the line of
sight is not an easy task.
The differences in the shape of the spectral features are hard to quantify systematically.
It is clear that this spectral variability correlates with the intrinsic color. 
In the literature a number of predictors have been suggested and studied with the aim of predicting the intrinsic color.
A simple method to study the color is to select a subsample of objects believed to have little or no reddening, and use them as a template \citep[e.g:][]{2005ApJ...624..532R,2010AJ....139..120F}.
More advanced methods involve the use of one or more spectral features as predictors of the intrinsic color \citep[e.g.][]{2011ApJ...734...42N,2011ApJ...742...89F,2011A&A...529L...4C}.
For example, the velocity of the \SiII~6355~\AA\ \citep{2011ApJ...729...55F,2014ApJ...797...75M}.
Finally, other methods are based on multi color light-curves   \citep[e.g][]{1996ApJ...473...88R,1996ApJ...473..588R,2010ApJS..190..418G}. The Lira-relationship is an example of such an approach \citep{1999AJ....118.1766P}
The properties of the extinction have also been studied with a statistical approach based on light curves \citep{2014ApJ...780...37S}.

{To extract the intrinsic color from the spectra is not a simple task.}
Partially, this is due to the difficulty in calibrating data obtained by slit spectroscopy.
 Even when flux calibrated spectra are available, it is difficult to separate the variability of intrinsic color
and luminosity from dust-extinction.
For a given spectral type, the intrinsic spectrum is the ``bluest'' in the
data. Extinction makes the spectra redder and fainter. However, when two
spectra are ``different'', the intrinsic color may be different.
The difficult part is quantify what ``different'' and ``similar'' mean when the spectra show a continuum distribution of properties.

Photometry can be used to construct light curves of SNe in different
bands, providing an indirect proxy of its spectral energy distribution. A fraction of the spectrum intrinsic variability is encoded in the photometry too. 
Most importantly, all the information 
about the extinction (type and amount of extinction) is 
encoded in the light curves. Also in this case it is difficult to
disentangle the intrinsic variability, which has to correlate with spectral
features, from the extrinsic variability due to dust extinction which affects the overall shape of the spectrum.

Our approach not only finds doubles of SNe with different extinction and similar spectal properties.
We use the reasonable assumption that when you have a SN with spectral characteristics that are intermediate between other two SNe, the intrinsic color is also intermediate between the two.
To find the function between the space of the spectra and the color curves we use a technique called Partial Least Square regression (PLS).
We use PLS to find the latent structures connecting
these two spaces and to predict the intrinsic component of the light curves
from series of spectra.
As shown by \cite{2015MNRAS.447.1247S} the data compression of the spectral
series is handled with the help of Expectation Maximization Principal Component Analysis (EMPCA) and derivative spectroscopy.

The paper is organized as follows. Section \ref{sec:sn_data} illustrates the SN data used in our analysis. Section \ref{sec:MultiPLS} explains the technique used. Section \ref{sec:results} shows the results on SN\,Ia data. Section \ref{sec:conclusions} wraps up the conclusions and future applications.

\section{The type Ia supernova data}
\label{sec:sn_data}

For this work we collected most of the currently publicly available visible
spectra and B and V band photometry of SNe\,Ia.  Our sources of spectra
 are the CfA spectroscopic release \citep{2012AJ....143..126B}, the
Berkleley Supernova Program \citep{2012MNRAS.425.1789S}, the Carnegie Supernova
Project \citep[CSP,][]{2013ApJ...773...53F}. We make use of the SN repositories SUSPECT
\footnote{\url{http://www.nhn.ou.edu/~suspect}} and WISEREP
\citep{2012PASP..124..668Y}.  The spectra are deredshifted using the
heliocentric redshifts published in \cite{2012AJ....143..126B}. CSP spectra are
published deredshifted.

We collect the $B$ and $V$ band photometry published in
\cite{2009ApJ...700..331H} and \cite{2011AJ....142..156S}.
The redshift of most of the sample is lower than $z=0.04$. The central wavelength of the filters in the observer frame is only marginally different from the corresponding wavelength in the rest frame.
 To check the importance of this effect, we performed the analysis with and without K-corrections on the photometry and in our sample the outcome is not significantly affected by this second order correction.
Of course, to study samples at high redshift, a careful K-correction becomes a necessity.
We check how important is to correct the photometry into rest frame before comparing our SNe\,Ia. We apply a K-correction that converts the magnitude to rest frame.
 As a first order approximation, we have chosen to use the spectra of SN\,2011fe \citep{2014MNRAS.439.1959M} as a template to calculate all of the K-corrections.
We shift the flux calibrated spectra of SN 2011fe to the redshift of the SN that we want to correct. 
Then, we calculate the K-correction factor at each epoch using the method from \cite{1968ApJ...154...21O,2002astro.ph.10394H},
and apply the relative K-correction to the photometric data at the nearest epoch. 
Applying K-corrections to our sample does not affect significantly our results, as it affects the fluxes in the studied bands by less than 5\% in most cases.

We deredden the photometry from the Milky Way extinction using \citep{ 1998ApJ...500..525S, 2011ApJ...737..103S}. The
host galaxy extinction, the subject of our study, is not estimated or removed from the photometry during this step.

To calculate the absolute magnitudes of the SNe we use CMB centered redshift measurements
from \cite{2009ApJ...700..331H} and assume a smooth Hubble flow. We are only
interested in differences between absolute magnitudes. This makes it unnecessary
to know the actual value of the Hubble constant.  The error on the absolute
magnitude due to the peculiar motion of each host galaxy is assumed to be 500
\kms\ \citep{2003MNRAS.346...78H}.

\section{The method: predicting light curves with spectral series}
\label{sec:MultiPLS}

This section summarizes the techniques used to construct a metric space for SN\,Ia spectral series \citep{2015MNRAS.447.1247S}
and to predict the intrinsic part of their light and color curves.


\subsection{Principal Component Analysis to construct a metric space for type Ia spectral series}

We follow closely the techniques explained in details in
\cite{2015MNRAS.447.1247S}.  In this subsection we recap how to create a metric
space for SN\,Ia spectral series.

In the literature a large number of visible spectra of SNe\,Ia
taken few weeks around maximum brightness is available.  Most of these spectra are
obtained by means of slit spectroscopy and many are not flux calibrated
\citep[e.g.][]{2012AJ....143..126B}. The spectra have been obtained with a number of
different instruments by many surveys over a time frame of decades. This means
that the wavelength coverage is not constant, the signal-to-noise ratio is 
diverse, and the time sampling of the SNe is not uniform.
The approach of \cite{2015MNRAS.447.1247S} offers a solution for these problems
and it creates a meaningful metric space for SN\,Ia spectral series.

We perform the following algorithms that lead to the creation of our metric space.  First, we calculate the
derivative over lambda of the spectra ($\frac{d \log_{10}F_{\lambda}}{d
\lambda}$).  The reason is to retain all the small scale information present in
the spectra, without the problem of the uncertainty on the distance and of the
 absolute flux calibration of the spectra.  Also, in
the derivative space, the weight given to long scale variations in the flux is reduced.  The
knowledge of the amount of extinction and the type of dust becomes unnecessary.
Most importantly, flux calibration of the spectra is also not necessary.

\begin{figure}
\centering
\includegraphics[width=1.\columnwidth]{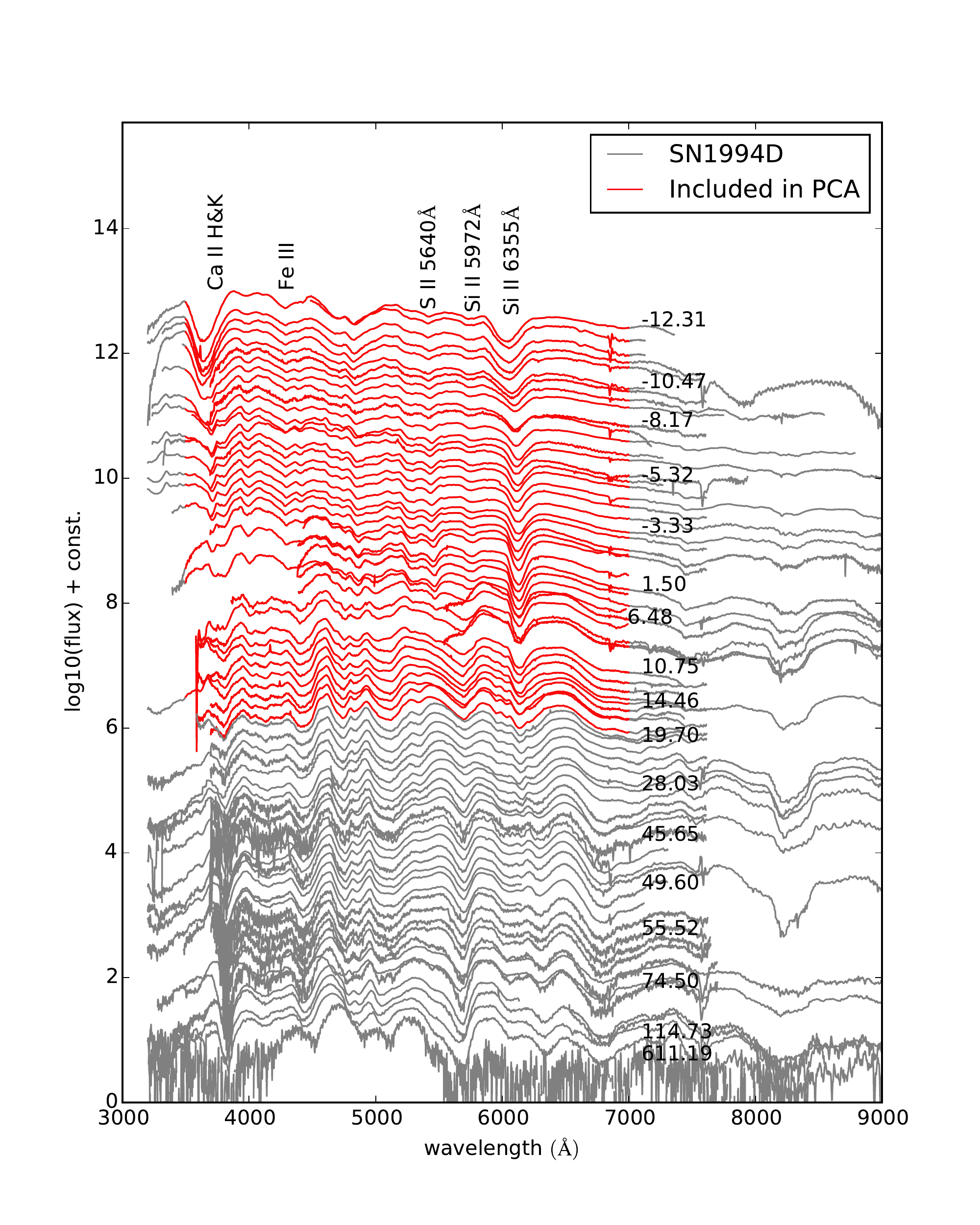}
\caption[Spectral evolution of a normal SN~Ia]{The figure shows the spectral
evolution of the prototypical normal SN~Ia, SN1994D. The evolution of the
spectrum was followed from two weeks before maximum up to few hundred days
after. The ``square'' delineated in red highlights the spectral variability
that we include in our statistical analysis.}
\label{fig:sn1994D_spectra}
\end{figure}

We limit the analysis in the 3500-7000 \AA\ range.
This is the wavelength range that is mostly complete in the data that we are using.
We consider only spectra
from $-12.5$ up to $+17.5$ days relative to $B$-maximum.
Spectral observations become scarce outside of this epoch range.
  This fraction of the
spectral variability of SNe\,Ia is highligted in Fig.~\ref{fig:sn1994D_spectra}
on the spectral series of SN\,1994D, a prototypical SN\,Ia.
The derivative spectra in the considered range of epochs are binned by time bins of $2.5$ days.
The input matrix of the analysis has a number of columns equal to the number of wavelength bins times the number of epoch bins, and a number of rows equal to the number of SNe.
Following \cite{2015MNRAS.447.1247S}, we apply EMPCA on this matrix to reduce the dimensionality of the data down to 5 principal components (PCs).


\subsection{Interpolating the photometry with Gaussian Processes}

 For
the Partial Least Square algorithm, we need to transform the
discrete time series of the photometry into a continuous function.
 We need light curves as a continuous
quantity with associated errors. These errors have to take into account the
uncertainty of the observed photometry and the sparsity of the data.
 We need a robust regression method to fit light curves and color curves.

{SN~Ia photometry is typically interpolated with the help of light
curve fitters.  These algorithms construct a parametrized template for SN~Ia
light curve.  When fitting only $B$ and $V$ bands, they usually employ only two
parameters \citep[e.g. SALT2][]{2007A&A...466...11G}.  The first parameter accounts for
the decline rate of the SN (e.g. \Deltam or stretch), the second parameter accounts for a color
correction using the observed color.  This implicitly assumes that SN~Ia light
curves are described by two parameters only. This is, of course, true only as a first
approximation, and from only the light curves  it is hard to extract many more
parameters.  But from the study of the spectra 
\citep{2005ApJ...623.1011B,2006PASP..118..560B,2006MNRAS.370..299H,2012AJ....143..126B}
 it is clear that SN~Ia are much
more diverse, and we want to take this diversity into account.  } For these
reasons we do not want to use typical SN light curve fitters. With these
fitters the result of the fit depends not only on the data of the SN under
consideration, but also on the rest of the sample. Of course this makes sense
if the aim is to calibrate the objects as well as possible, but it is not the
best approach if we want to study the relations between light curves and
spectra in an unbiased way. 

A simple approach could be to obtain the light curves with the least square fit
to a polynomial. This introduces biases in the result. Arbitrary choices such
as the degree of the polynomial will affect the outcome.
A better approach is the use of a spline fit. This is equivalent to using a
series of polynomials smoothly connected together, but also this approach is
inadequate for our purposes because it does not give a time dependent
uncertainty on the result of the fit.

\begin{figure}
\centering
\includegraphics[width=1.\columnwidth]{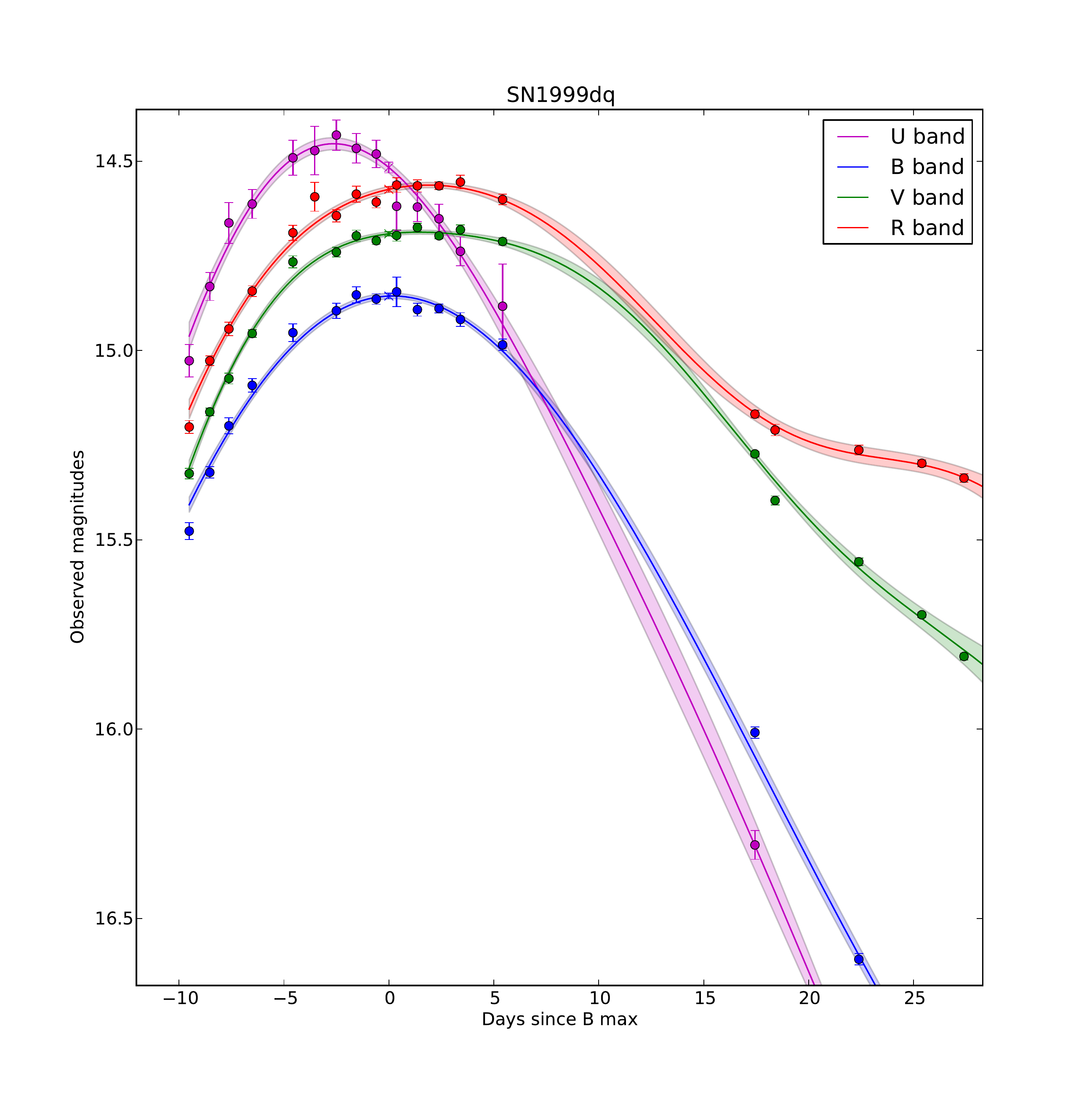}
\caption[Gaussian Processes]{ Light curve fitting in four bands of the SN\,1999dq.
The fit is the result of a Gaussian Processes regression.
}
\label{fig:gp_lc}
\end{figure}

We use a powerful regression method based on Gaussian Processes
\citep{Rasmussen:2005:GPM:1162254, 2012PhRvD..85l3530S, 2013ApJ...766...84K}. Gaussian Processes Regression is a more
general approach to determine the underlining function from a sparse set of
data (Appendix \ref{Appendix}).
As an example, the result of Gaussian Process Regression on the photometry of the SN~1999dq
is shown in Fig.~\ref{fig:gp_lc}. Now we have robust interpolated luminosities
with sensible error estimates for the fitted light curves. Most importantly,
the fits are independent from the behaviour of the other SNe of the
sample.

\subsection{Multivariate Partial Least Square}
\label{sec:PLS}

We want to predict the light curves of SNe~Ia from the PC space on the time
series of spectra. A good
tool for this job is Partial Least Square regression \citep{wold2001pls}.
\cite{2015MNRAS.447.1247S} use 
univariate PLS, here we use the
multivariate version of this regression technique.

The goal of PLS regression is to predict the $\mathcal{Y}$ space of the responses from the
$\mathcal{X}$ space of the predictors to recover their common structure. PLS finds the linear
relations that allow to predict a set of quantities in the $\mathcal{Y}$ space,
called responses, from the space of predictors $\mathcal{X}$. The underlying
assumption is that every component of the space of responses is a linear
combination of the predictors.
Mathematically the relation can be written as:
\begin{equation}
\textsf{\textbf{X}} = \textsf{\textbf{T}} \textsf{\textbf{P}}^T + residuals 
\end{equation}
\vspace{-.9cm}
\[\textsf{\textbf{Y}} = \textsf{\textbf{U}} \textsf{\textbf{Q}}^T + residuals\]
where $\textsf{\textbf{X}}$ is the training set of predictors.
$\textsf{\textbf{X}}$ has dimensions $N\times M$ where $N$ is the number of
observations (the number of SNe) and $M$ is the dimension of the predictor
space $\mathcal{X}$. In our case $M$ is the dimension of the PCA space built from
spectra. The matrix $\textsf{\textbf{Y}}$ is the matrix of the training set of
responses. It has dimensions $N\times L$ where $L$ is the number of epochs
sampled from the light curve. \textsf{\textbf{T}} and \textsf{\textbf{U}} have
dimensions $M\times n$ where $n$ is the dimensionality reduction of PLS, that
is the number of components chosen to explain the relation between the space
$\mathcal{X}$ and the space $\mathcal{Y}$.
$\textsf{\textbf{P}}$ and $\textsf{\textbf{Q}}$ are called loadings.
The decomposition is made so as to maximize the covariance between
$\textsf{\textbf{P}}$ and $\textsf{\textbf{Q}}$. The loadings become good
summaries of the spaces $\mathcal{X}$ and $\mathcal{Y}$ and the residuals
become ``small''.
$\textsf{\textbf{T}}$ represents the projections on the latent structures
defined by the matrix of the weights $\textsf{\textbf{W}}$:
\begin{equation}
\textsf{\textbf{T}} = \textsf{\textbf{W}} \textsf{\textbf{X}}. 
\label{eq:compression}
\end{equation}

The scores ($\textsf{\textbf{T}}$ and $\textsf{\textbf{U}}$) have the property
that they reproduce well $\textsf{\textbf{X}}$ and $\textsf{\textbf{Y}}$, and
the x-scores ($\textsf{\textbf{T}}$) are good predictors of
$\textsf{\textbf{Y}}$:
\begin{equation}
 \textsf{\textbf{Y}} = \textsf{\textbf{T}}
\textsf{\textbf{Q}}^T + residuals,
\label{eq:prediction}
\end{equation}
 when the residuals of the prediction have to be
small.

The relation between the two spaces is best explained in Fig.~1
 in \cite{wold2001pls}.  The dimensionality of the initial
space gets reduced to $n$, the dimension of the space of the latent variables,
by the matrix $\textsf{\textbf{W}}$ (equation \ref{eq:compression}). 
  Then, the matrix $\textsf{\textbf{T}}$ is
responsible for predicting the space $\mathcal{Y}$ trough equation
\ref{eq:prediction}.  In our case, the ``structure descriptors'' are the
spectral series and the ``activity measures'' are the light curves. The
variables are the coefficients of the PCA space of the spectra, the
observations are the different SNe, \{t$_1$, t$_2$, t$_3$\} are the latent
variables, $\textsf{\textbf{Y}}$ is the matrix of the observed light curves, M
is the dimension of the range of epochs included in the light curves.

Multivariate PLS is particularly recommended when there is a high correlation
between the responses. This is the case for the magnitudes at different epochs of light curves and color curves.

The decompositions of \textsf{\textbf{X}} and \textsf{\textbf{Y}} are chosen to
explain as much as possible of the covariance between the two datasets.

A simple algorithm to compute the weight matrix (\textsf{\textbf{W}}) and the
scores (\textsf{\textbf{T}} and \textsf{\textbf{U}}) proceeds as shown in
Algorithm \ref{algo:PLS}.
This algorithm is implemented in \textit{scikit-learn}
\citep{scikit-learn}.

The tools described before were made publicly available in the \href{https://github.com/sasdelli/lc_predictor}{LC\_predictor} Python package\footnote{\url{https://github.com/sasdelli/lc_predictor}}.
The code takes as input the series of spectra of a SN\,Ia and automatically performs the smoothing, and it projects the data on the trained PCA and PLS spaces.
It outputs the reconstructed spectral series as a sanity check, and it produces the predicted light curves and color curves after reddening corrections.

\begin{algorithm}
\caption{Partial Least Square algorithm}
\begin{enumerate}

\item Assign $\textsf{\textbf{X}}_0=\textsf{\textbf{X}}$ and $\textsf{\textbf{Y}}_0=Y$ (first iteration)

\item repeat $n$ times (the chosen dimensionality reduction) 

\ \ \ \ \ \ Compute the Singular Value Decomposition (SVD) of the matrix $\textsf{\textbf{X}}_n^T \textsf{\textbf{Y}}_n$ 

\ \ \ \ \ \ Compute the first left singular vector ($w_n$) of the matrix $\textsf{\textbf{X}}_n^T Y_n$. 

\ \ \ \ \ \ Compute the first right singular vectors ($v_n$) of the matrix $\textsf{\textbf{X}}_n^T \textsf{\textbf{Y}}_n$. 

\ \ \ \ \ \ Compute the $n$th X-score: $T_n$ = $\textsf{\textbf{X}}_n w_n $

\ \ \ \ \ \ Compute the $n$th Y-score: $U_n$ = $\textsf{\textbf{Y}}_n v_n $

\ \ \ \ \ \ Deflate the $\textsf{\textbf{X}}$ matrix: $\textsf{\textbf{X}}_n+1$ = $\textsf{\textbf{X}}_{n} - T_{n} P_{n}^T $. It is not necessary to deflate $\textsf{\textbf{Y}}$.

\end{enumerate}
\label{algo:PLS}
\end{algorithm}


\section{Predicting Light Curves and Color Curves from the Spectra}
\label{sec:results}

In this section we use multivariate PLS regression to find correlations
between photometry and spectral properties. The spectral
properties are described by the coefficients of the PCA space constructed from
spectral series. The coefficients from the PCA
encode the variance within Type Ia SNe spectral series in 5 components.
 The good quality of the reconstructions of the spectra prove that
this decomposition encompasses the variability of Type Ia spectra. This
space, by construction, does not include reddening.

The intrinsic colors and absolute magnitudes are a function of the spectral
series. Physically, a given spectral series is expected to have a corresponding intrinsic color curve and light curve. That is, the luminosity and
color at a certain epoch are expected to be a function of the components of the PCA space constructed from the
spectra. We use PLS regression to extract this function. Using this approach,
we make the implicit assumption of linearity. That is, the intrinsic colors and
absolute magnitudes are assumed to be well approximated by linear functions of the components of
the PCA space of the spectra.

The observed colors are subject to reddening.  To find the intrinsic color one
has to find the locus of the bluest SN for every point of the PCA space.  We
use an approach similar to what was used by \cite{2015MNRAS.447.1247S} to discard SNe
with large reddening.  In that case we had only one parameter for the
treshold of maximum reddening to select the SNe, here we have more complicated
color curves and light curves.
  Additionally, we are using photometric data that have a larger diversity in their errors.
  We select
supernovae for the PLS regression if they have a reddening lower than a given
treshold above the PLS prediction and
with errors lower than a given value in all the epochs.  The
PLS algorithm and the selection is run a number of times until the solution
has converged \citep[see section 5.4 of ][]{2015MNRAS.447.1247S}.  The supernovae in each iteration can be selected or deselected.
These parameters are validated through cross-validation.

\subsection{Predicting the Light Curve from the Spectra}

We apply the approach to the $B$-band light curves between $-5$ days and $+35$
days from maximum. The aim is to predict the absolute magnitude curve using the
spectra of the supernova. This is done free from assumptions on reddening laws
and extinction. First of all the observed magnitudes need to be scaled to the
reference frame. This is simply achieved using the host-galaxy redshift and
assuming a smooth Hubble flow. The observed magnitude needs to be decreased by the
distance modulus:
\[\text{M}_{abs} = \text{M}_{obs} -\mu\]
where, assuming an homologous expansion for the Hubble flow:
\[ \mu = 5(\log_{10}( z c / H{_0} )+5).\]
 The peculiar motion of the galaxies will add an error to this estimate.
 The error on the absolute magnitudes due to the
dispersion of the peculiar velocity $\sigma(v)$ of the galaxies becomes:
\[\sigma_\mu = 5 \sigma(v) /( \ln(10)z c).\]

\begin{figure*} 
\centering
\includegraphics[width=1.\textwidth]{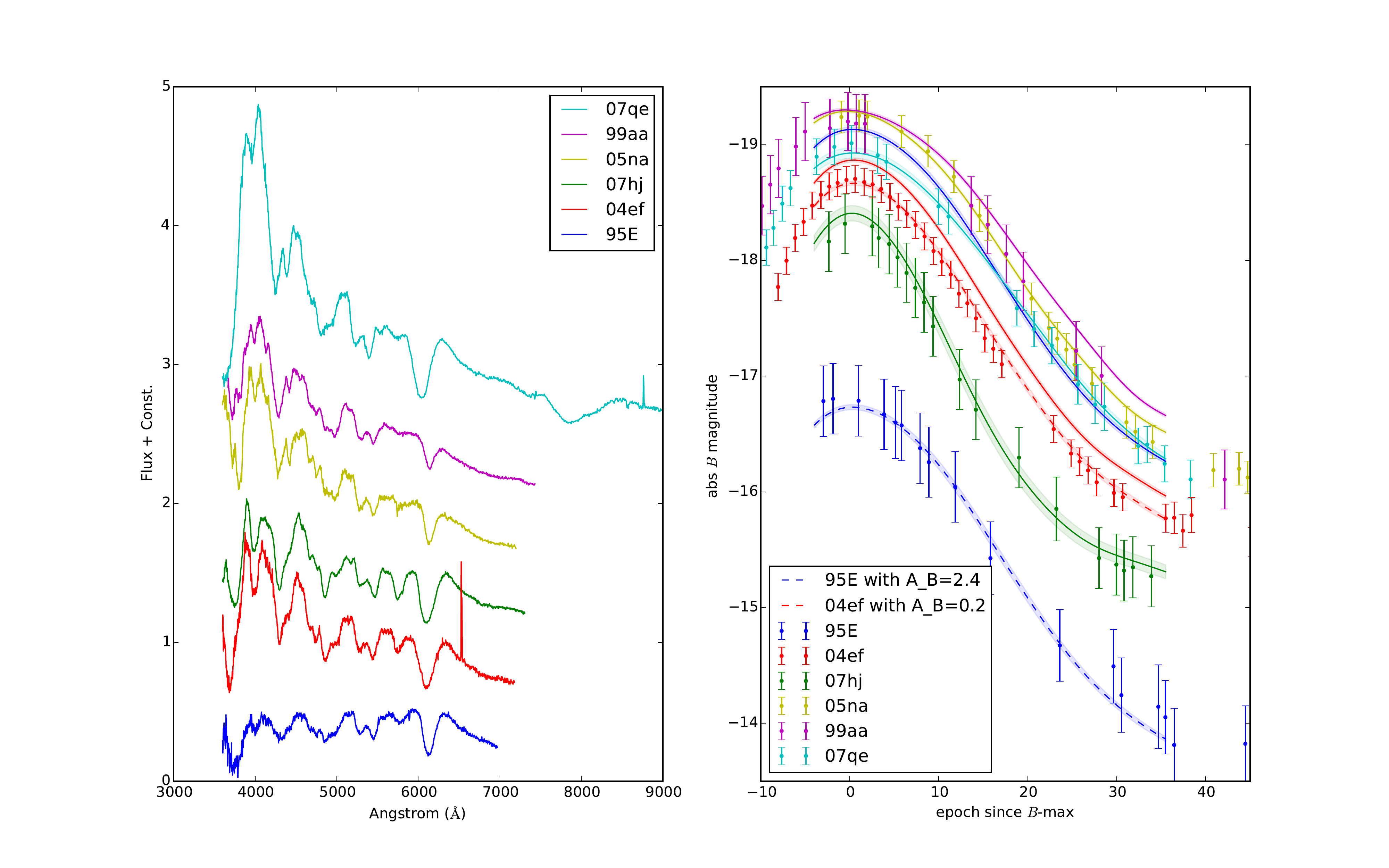} 
\caption[$B$-band light curve predicted by the spectra]{
the left and the right panel represent the space of predictors and the space of responses and correspond to 
the X and Y spaces of section \ref{sec:PLS}. 
The left panel shows the spectra close to B maximum of some SNe examples. The fluxes are in arbitrary units and the spectra have been shifted for clarity.
The right panel shows the photometry of the same SNe.  The solid curves are the predictions based on PLS regression on the PCA space on the spectra.  
The blue dashed curve show the effect of adding extinction on the predicted light curve of the SN\,1995E.}
\label{fig:Bmag_lc_pred}
\end{figure*}

The left panel of Fig.~\ref{fig:Bmag_lc_pred} shows, as an example, the spectra of a few SNe close to maximum brightness. The significant variability in spectral properties present in the data is apparent from this selection.
In the right panel we show the corresponding light curves of these SNe.
 The points are the original photometry, corrected for Hubble
flow, but with no attempts to correct for reddening.  The solid curves show the
corresponding PLS predictions. The colored area represents the uncertainty of
the prediction calculated using k-folding.  The photometric data and the
predictions match nicely.  SN~1995E, however, has a luminosity
significantly lower than is predicted.  This is expected and it can be explained by a significant amount of reddening of
this individual object.
  As it was shown in \cite{2015MNRAS.447.1247S}, reddening
does not influence the components of the PCA space.
Consequently, a  supernova with high
reddening does not have a PLS prediction different from a supernova with no
reddening.
  From this we deduce that the mismatch between the prediction and
the observations comes from dust extinction.
On individual objects the mismatch can be slightly negative.
However, these mismatches are always consistent with zero, i.e. no reddening.
This is due to random uncertainties in the observed data, in the distance estimates and in the analysis.

\begin{figure} 
\centering
\includegraphics[width=1.\columnwidth]{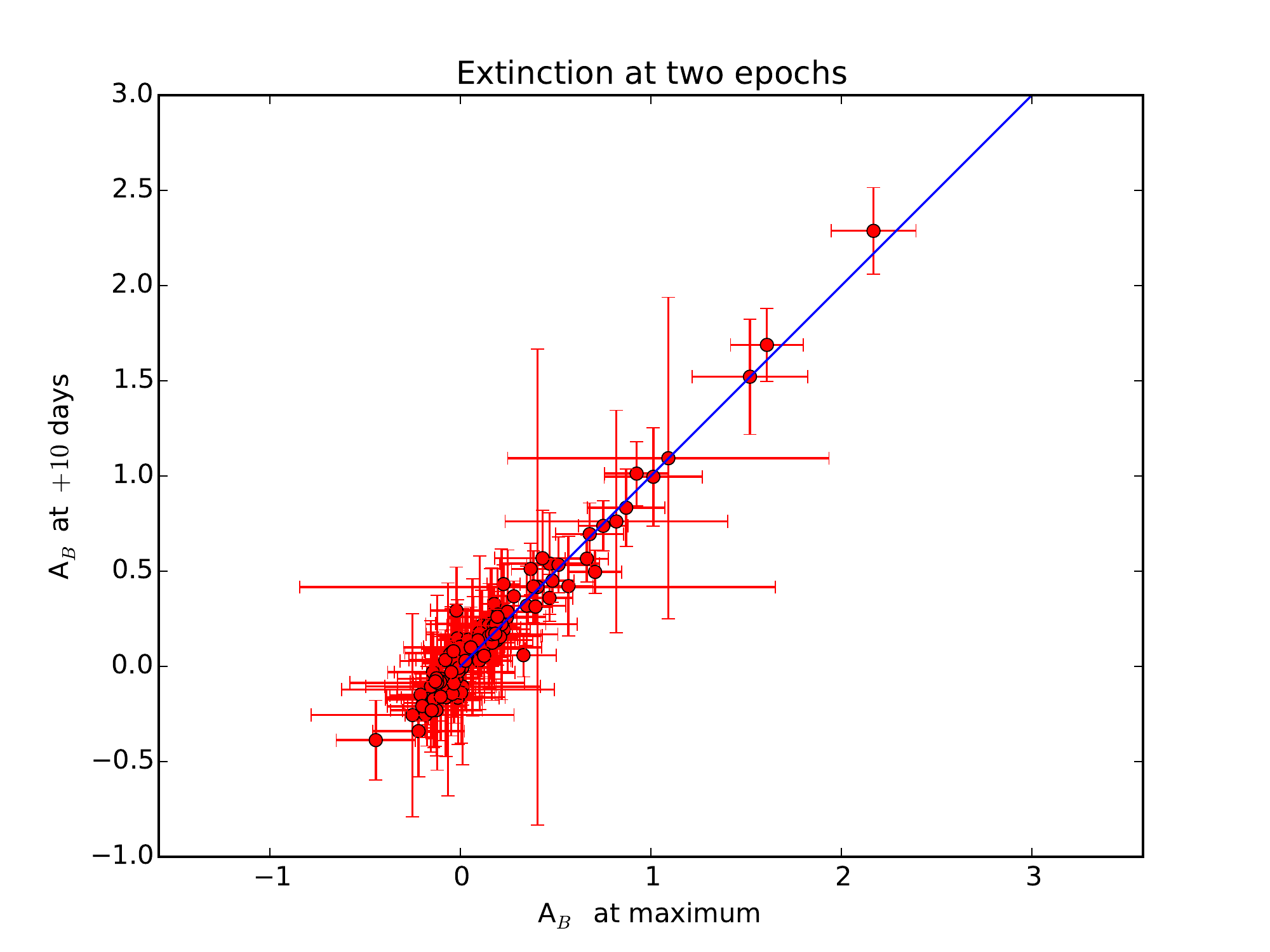} 
\caption[Dust extinction in the $B$-band]{The figure shows the A$_B$ at maximum and at $+10$ days obtained by 
PLS regression. The solid line is not a fit. It is just a diagonal overplotted as a reference. }
\label{fig:EBmag_at_two_epochs} 
\end{figure}

\begin{figure} 
\centering
\includegraphics[width=1.\columnwidth]{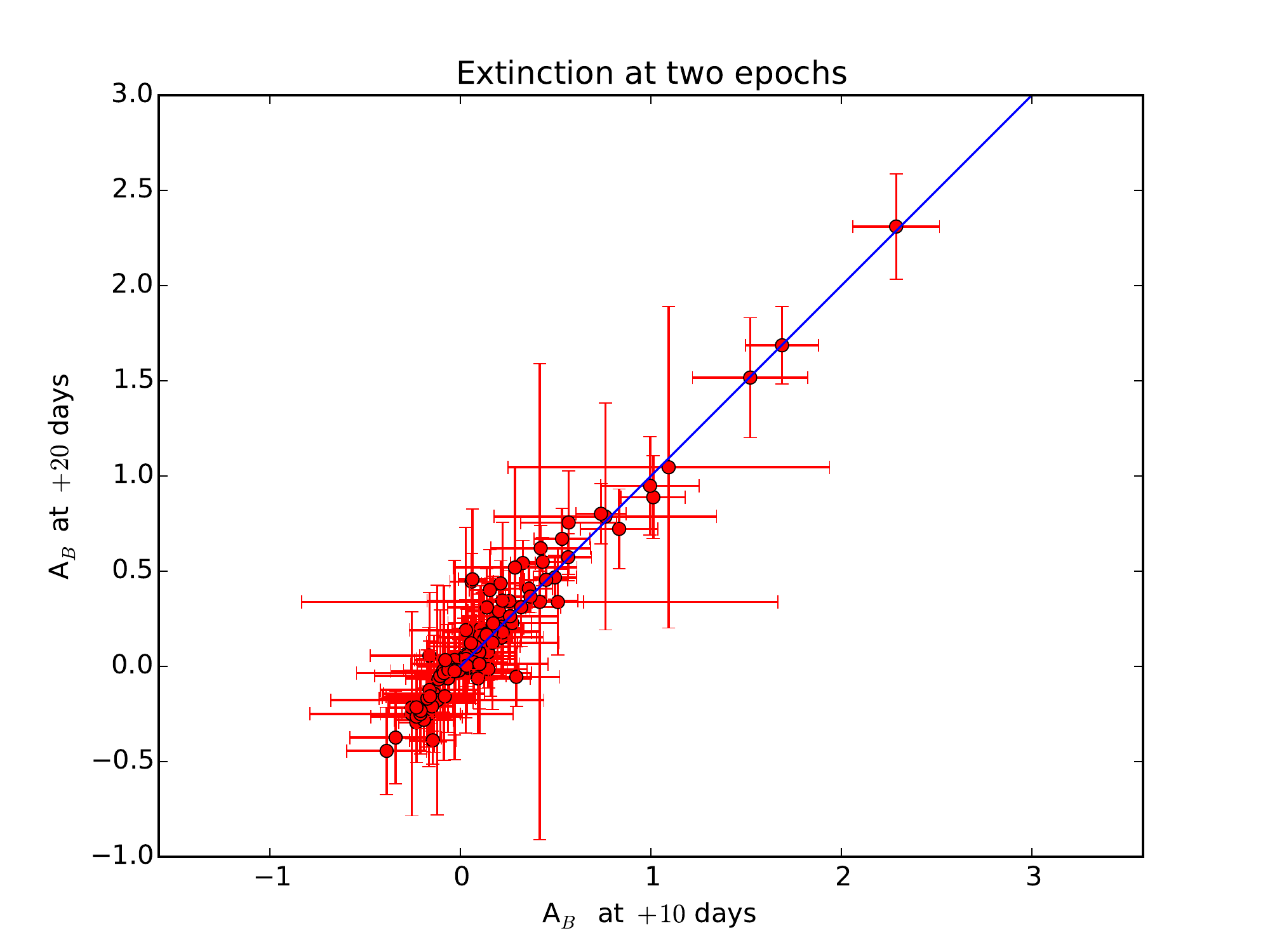} 
\caption[Dust extinction in the $B$-band]{The figure shows the A$_B$ at $+10$ days and at $+20$ days obtained by 
PLS regression. }
\label{fig:EBmag_at_two_epochs_2} 
\end{figure}

  The difference between the curve
and the data is an estimate of the extinction independent from assumptions on the nature
of the reddening law or the amount of reddening. Clearly, this estimate can be calculated at
different epochs.  Under the assumption that the amount of extinction does not
vary with time, the luminosity deficit of a supernova in a given band as a first order approximation stays constant.
  We check it in Fig.~\ref{fig:EBmag_at_two_epochs}.
  It shows the
extinction in the $B$ band (A$_B$) at maximum and at $+10$ days for all the
supernovae in our sample.
Fig.~\ref{fig:EBmag_at_two_epochs_2} shows the extinction in the $B$ band at $+10$ days and at $+20$ days.
The extinctions are consistent between each other, which implies that the extinction
does not vary significantly after maximum, and supports the reliability of
our method.

%

\subsection{The Phillips-Relation}

\begin{figure}
\centering
\includegraphics[width=1.\columnwidth]{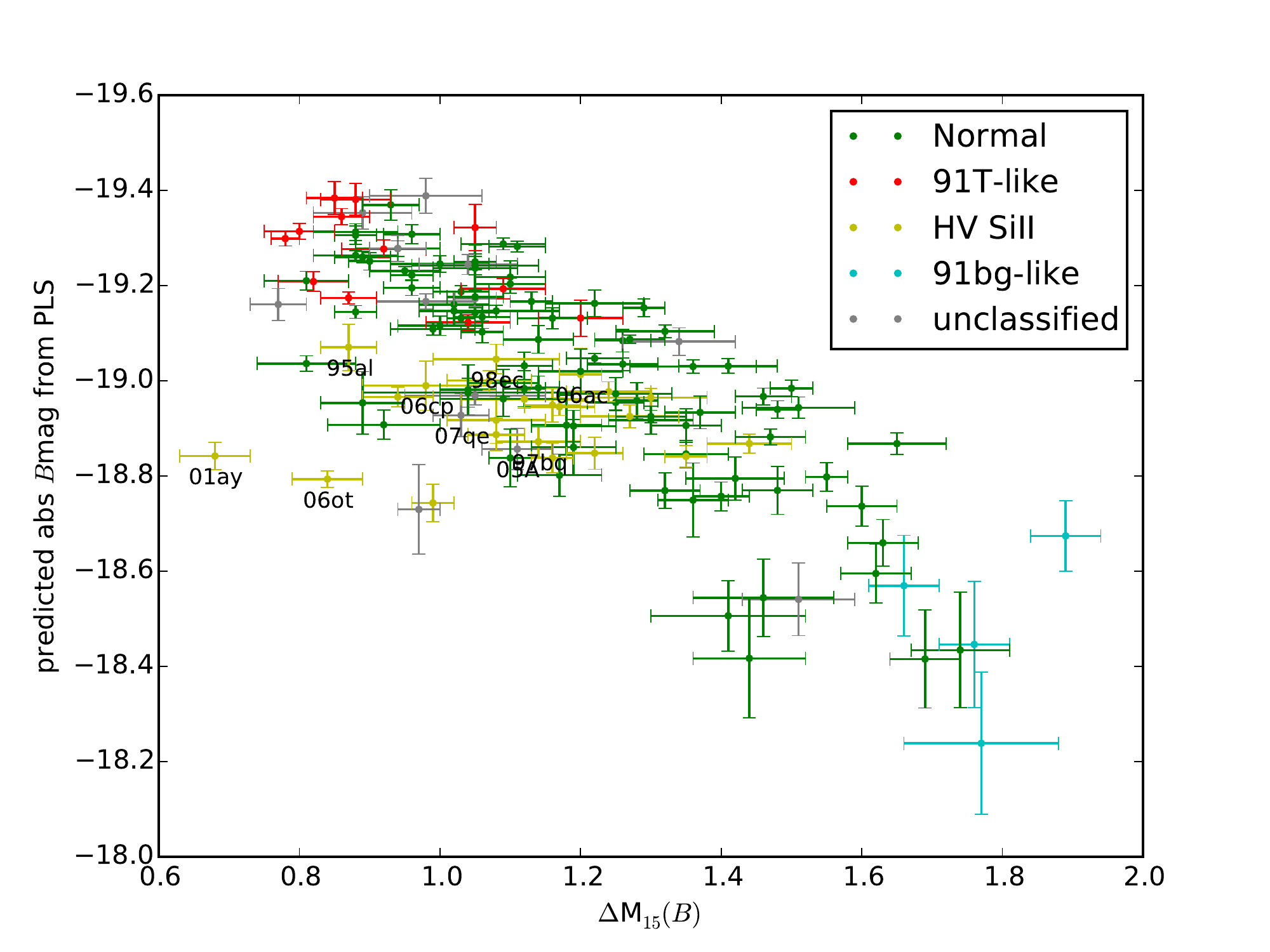}
\caption[The Phillips relationship]{ The Phillips relationship between the $B$-band luminosity at maximum
predicted by PLS and the decline rate of the light curve. The different subclasses
of SN~Ia are colored \citep[][]{2001ApJ...546..734L,2009ApJ...699L.139W}. The SNe that are close to SN~2001ay in the PCA space are tagged.}
\label{fig:Phillips_rel}
\end{figure}

Type Ia SNe luminosity is known to anti-correlate with the decline in
luminosity after maximum \citep{1993ApJ...413L.105P}.
 From the PLS
regression we have an estimate for the $B$-band peak luminosity independent from
reddening assumptions.
  In Fig.~\ref{fig:Phillips_rel} we show the relation
between our estimate and the \Deltam, the difference between the magnitude at maximum and
at $+15$ days.  Many of the known characteristics of SN~Ia show up in this
diagram. On the bottom right are the faint and fast declining 1991bg-like SNe.
There are only a few objects of that kind in our sample, hence the errors on the
predicted magnitudes are large.
  Spectroscopically normal SNe show a wide range
of luminosities and decline rates.
  On the tip of the relation are the luminous
1991T-like SNe.
  An interesting ``outlier'' of the Phillips relationship is the
SN~2001ay.
 It is the left-most point in the figure, with a \Deltam = 0.7 (the
slowest of the sample).
  This SN is so extreme that is was clearly recognized
as an outlier of the relation by \cite{2011AJ....142...74K}.
 It is too faint for its decline rate.
Our analysis nicely confirms it and shows that such outliers (although less extreme) are not uncommon.
This suggests that there is a population of SNe with characteristics that are
intermediate between SN~2001ay and the bulk of SNe\,Ia. These SNe are highlited in Fig.~\ref{fig:Phillips_rel}. They can
be roughly identified with the SNe~Ia with a high photospheric velocities in
their lines \citep[HV in the classification scheme from][]{2009ApJ...699L.139W}.
Many of these SNe are ``underluminous'' compared to what their \Deltam\
predicts. Most of these SNe are characterized by high photospheric velocities
and intrinsically redder colors than the average. 

\subsection{Predicting the Color Curve from the Spectra}

In this section we apply the PLS regression method on $B-V$ color curves.
Differently from magnitudes, the observed colors are not affected by the distance, and
can usually be measured precisely in nearby SNe.  However, the
intrinsic variance of colors is smaller than the variance in the magnitudes.
This makes the regression task quite complicated.

\begin{figure}
\centering
\includegraphics[width=1.\columnwidth]{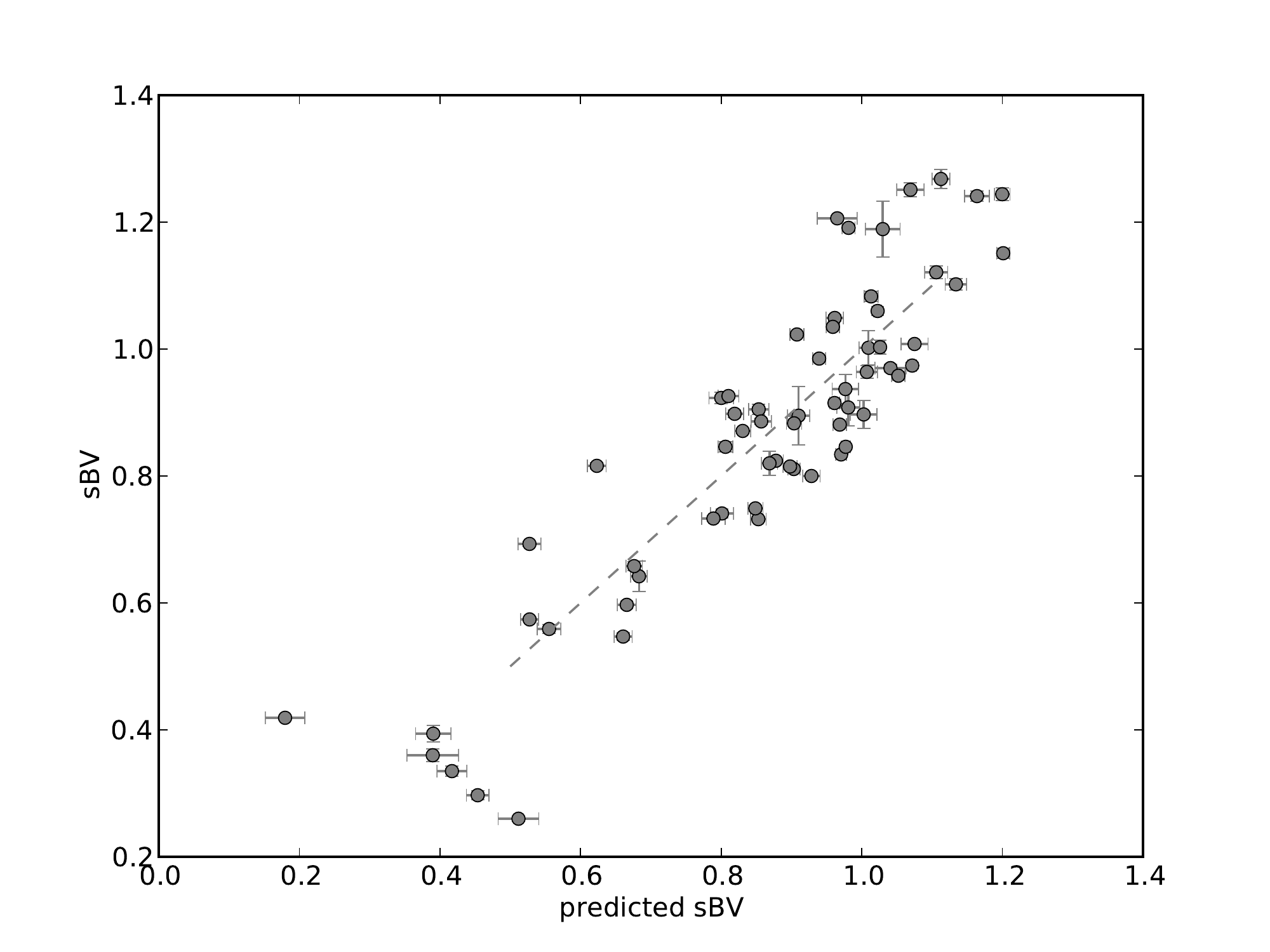}
\caption[The epoch of the $B-V$ color maximum]{ The relation between the epoch of the $B-V$ maximum in units of $30$
days \citep[s$_{BV}$,][]{2014ApJ...789...32B} and its prediction from PLS regression performed on the PCA space of spectral
properties.}
\label{fig:sbv}
\end{figure}

First, we want to show that important characteristics of the $B-V$ color curve
are retained in the PCA space of spectra.  The time between the $B$ maximum
epoch and the maximum in the $B-V$ color curve is an almost reddening independent
quantity. Its use was suggested for SN~Ia classification and cosmology \citep{2014ApJ...789...32B} as an alternative to \Deltam. The
maximum in the $B-V$ color curve happens usually at about $+30$ days after maximum.  This quantity 
is shown to correlate with the spectral properties encoded in 
our PCA space. Simple univariate PLS regression between the PCA space and this 
color curve indicator shows an excellent correlation (Fig.~\ref{fig:sbv}).

\begin{figure*} 
\centering
\includegraphics[width=1.\textwidth]{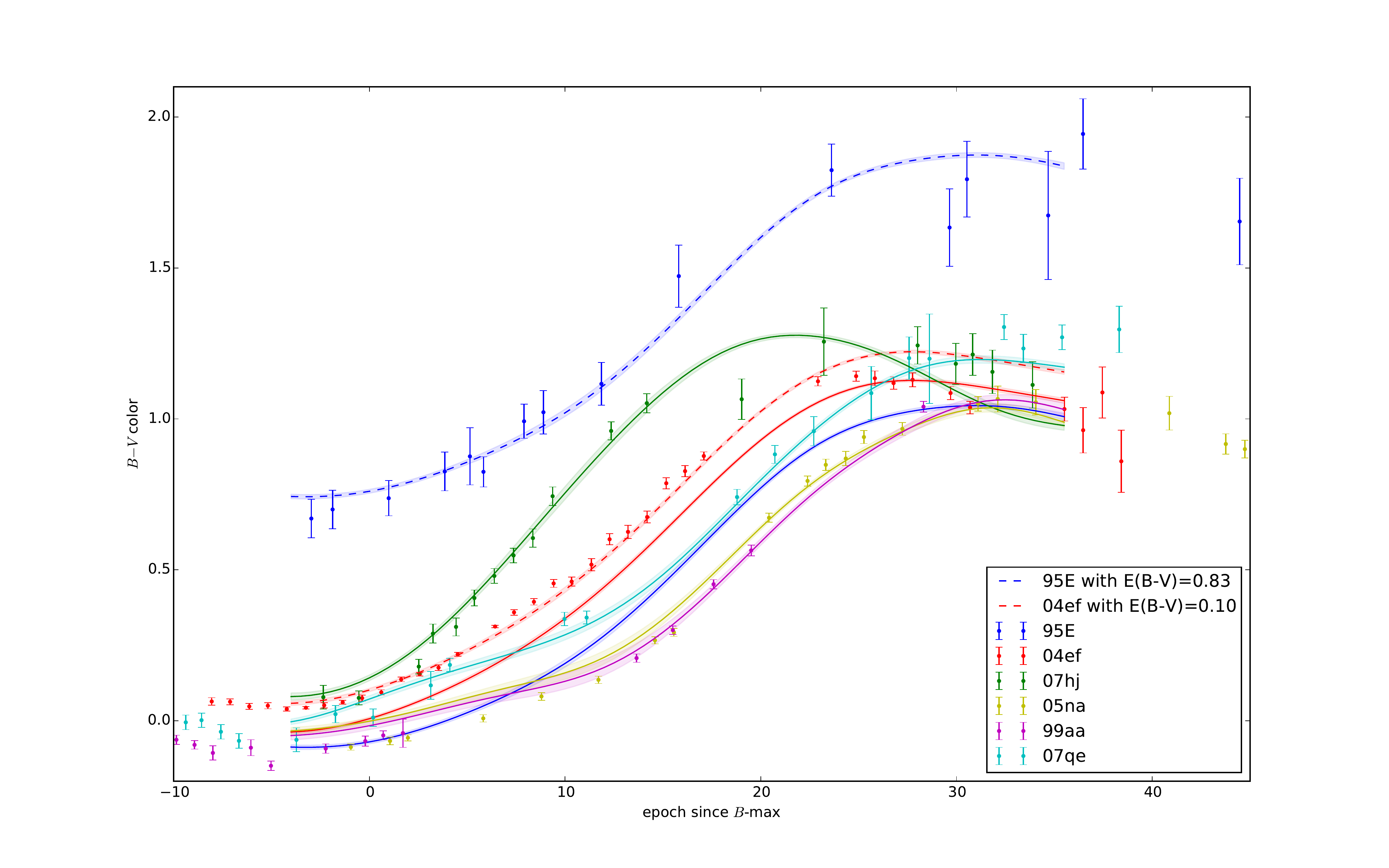}
\caption[$B-V$ color curve predicted by the spectra]{
The figure shows the observed $B-V$ colors for a number of SNe (dots with errorbars).
These are compared with the color curves predicted
from the spectra by means of PLS regression (solid lines).
The blue dashed curve shows the effect of adding extinction on the predicted color curve of the extinguished SN\,1995E.
}
\label{fig:BmV_lc_pred}
\end{figure*}

\cite{2015MNRAS.447.1247S} also showed that the color at maximum has a significant correlation with 
spectral properties.
Here we generalize these individual approaches with the help of multivariate PLS.
Fig.~\ref{fig:BmV_lc_pred} shows some observed color curves (points with
errorbars) and the corresponding predictions from the spectra using PLS (solid lines).
The left panel of Fig.~\ref{fig:Bmag_lc_pred} shows the spectra of the corresponding SNe at maximum.
It is evident that the colors are quite uniform at maximum, later they have a
large spread of properties.
SN~1995E has clearly a significant color excess in comparison with the
prediction from its spectra.
This is consistent with a significant amount of reddening.
The analysis converts the spectral variability present in the spectra
 into a prediction for the intrinsic color curve (solid curve).

\begin{figure} 
\centering
\includegraphics[width=1.\columnwidth]{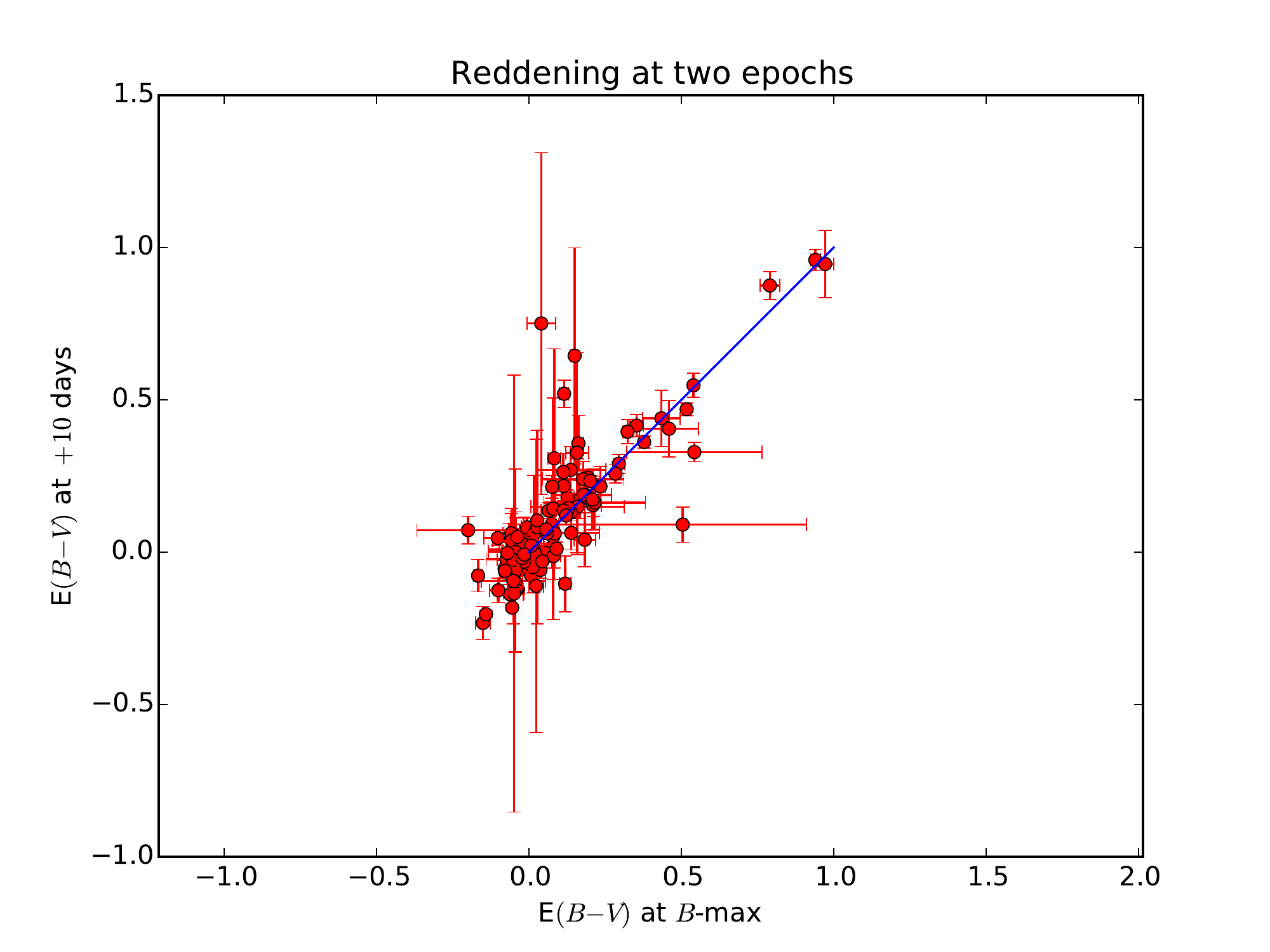} 
\caption[Dust extinction of the $B-V$ color]{ The figure shows the $E(B-V)$ predicted by mulivariate PLS regression at maximum and at $+10$ days. A diagonal is overplotted as a reference. }
\label{fig:EBmV_at_two_epochs} 
\end{figure}

\begin{figure} 
\centering
\includegraphics[width=1.\columnwidth]{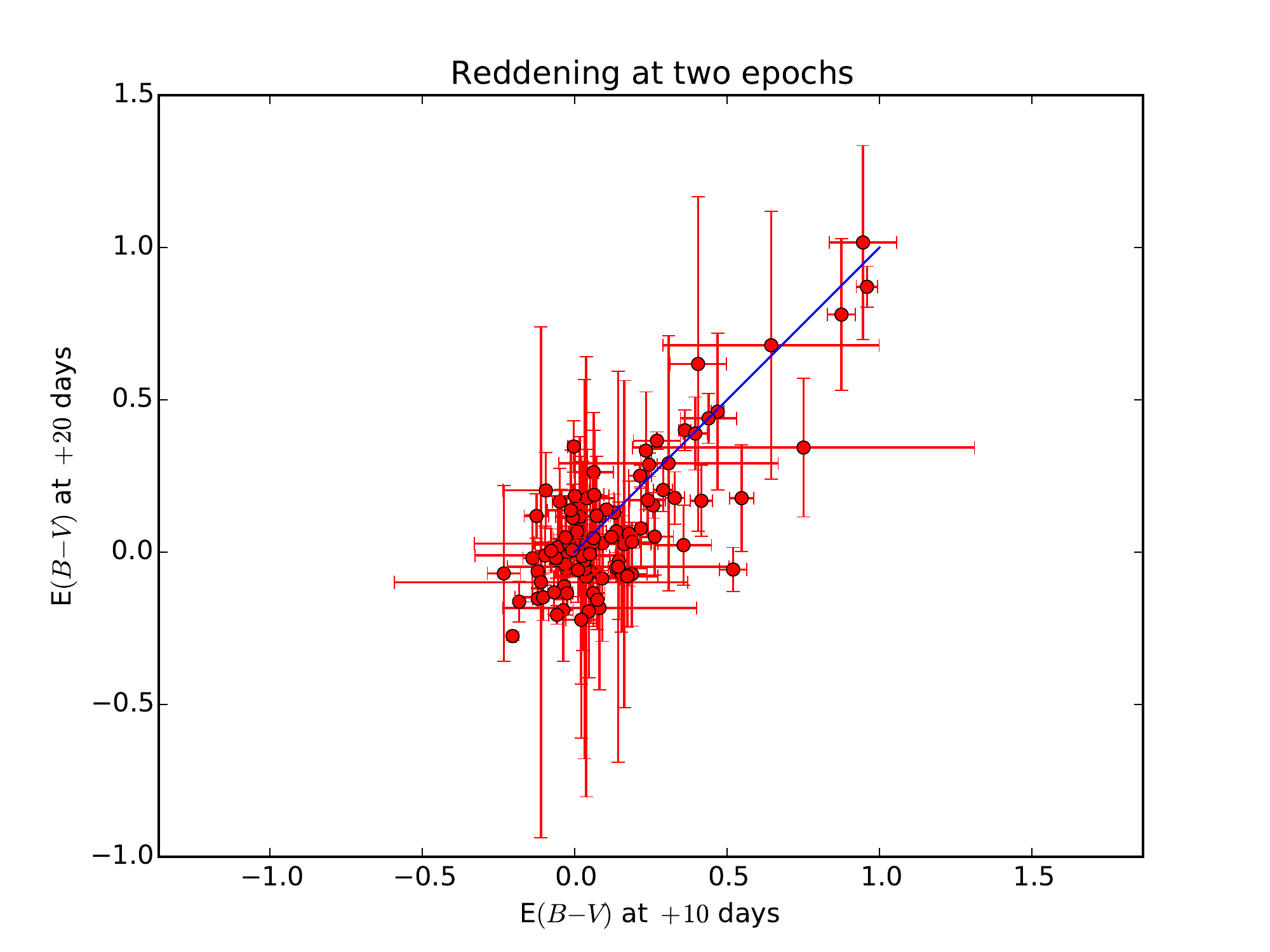} 
\caption[Dust extinction of the $B-V$ color]{ The figure shows the $E(B-V)$ predicted by mulivariate PLS regression at $+10$ and at $+20$ days. }
\label{fig:EBmV_at_two_epochs_2} 
\end{figure}

Similarly to before, we want to check that the PLS regression is predicting
most of the intrinsic color variability.
Under the assumption that the amount of reddening does not vary significantly
with time, we 
check that the color excess attributed to dust is constant at different epochs.
Figs.~\ref{fig:EBmV_at_two_epochs} and \ref{fig:EBmV_at_two_epochs_2} compare
the color excess at maximum, at $+10$ and at $+20$ days.
The agreement between the quantities supports that the regression 
predicts the majority of the intrinsic color variation and evolution and that the 
amount of reddening does not change significantly in this range of epochs.

%

\subsection{The Extinction-Reddening Relation}

The extinction due to dust affects most short wavelengths.  This means that
the dimming in the $B$ band (A$_B$) will be larger than the
effect in the $V$ band.
Hence, the relation between the extinction in these bands is usually 
parametrized as:
\[A_V = R_VE(B-V).\]
Typical values for $R_V$ measured in our galaxy vary between $2.1$ and $5.8$
with $3.1$ being the most common value \citep{1989ApJ...345..245C,2003ARA&A..41..241D}.

\begin{figure} 
\centering
\includegraphics[width=.88\columnwidth]{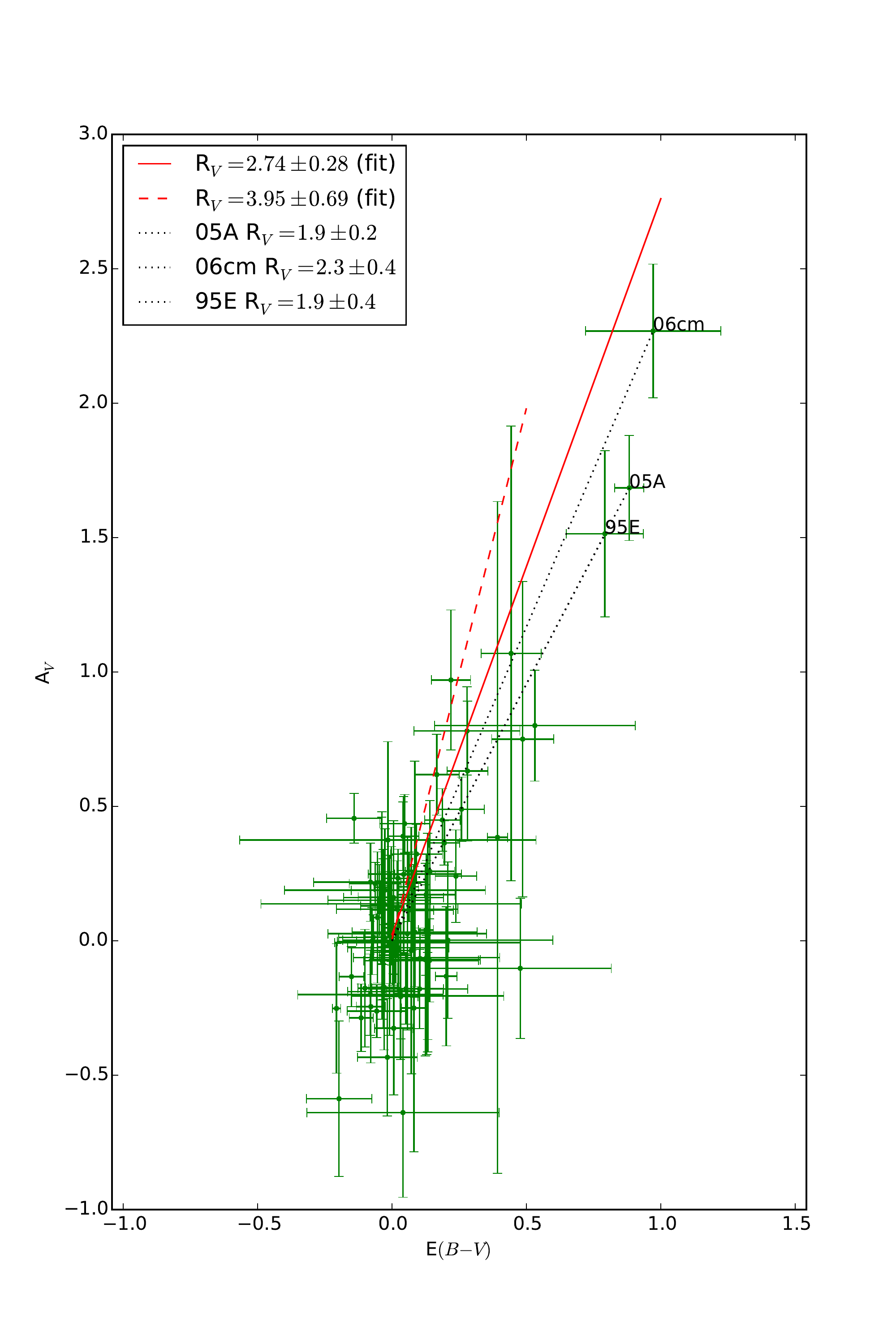}
\caption[Reddening law]{The plot shows the relation between the host-galaxy extinction in $V$ and 
reddening in $B-V$.  The solid line shows the result of a fit of the
relation. 
The dashed line shows the fit excluding the three most reddened
SNe, for which an individual estimate of $R_V$ is calculated.
The errors come from the measurement errors and from statistical errors
(k-folding) added in quadrature.
}
\label{fig:reddening_law} \end{figure}

Fig.~\ref{fig:reddening_law} shows the color and magnitudes excess from the 
PLS reconstruction.
  Frequently SN~Ia are associated with a low $R_V$ (frequently $<2$)
that does not show up in galactic reddening
\citep[e.g.][]{2007ApJ...664L..13C,2011ApJ...731..120M, 2014ApJ...789...32B}. This have been
explained by peculiar environments around SN~Ia
\citep[e.g.][]{2008ApJ...686L.103G, 2014MNRAS.443.2887F}.
In contrast, techniques based on spectral features return values of $R_V$
much more similar to what is typical in the Milky Way
\citep{2011A&A...529L...4C}.
Assuming that a significant fraction of the residuals in the 
Hubble diagram is due to intrinsic color instead of the intrinsic
 luminosity also favours a Milky-Way R$_V$ \citep{2014ApJ...780...37S}.
Performing an orthogonal distance regression
fit on the results of our analysis returns an $R_V =
2.78\pm0.28$ (Fig.~\ref{fig:reddening_law}). 
  This shows that the average host extinction of SNe\,Ia is
  consistent with the typical extinction
laws of the Milky-Way (R$_V=3.1$).
 In Fig.~\ref{fig:reddening_law} we also perform a fit excluding the three most
extinguished SNe of our sample. This confirms what was found by
\cite{2010AJ....139..120F}. SNe with an high extinction often have low values
of $R_V$ (Fig. \ref{fig:reddening_law}), and SNe with a low extinction show an
average R$_V$ similar to typical Milky Way values (dashed line).

It is hard to discriminate between
spectroscopically different SNe by using only light curves.
Certain spectral types (mostly SNe with high photospheric velocities) have light curve and color curve
shapes similar to average SN\,Ia but fainter and redder.
Techniques of color-curve matching may find a match between two SNe that are intrinsically different and think that differences that are intrinsic dome from dust extinction. 
 The ratio
between this ``missing luminosity'' and ``color excess'' happens to be
significantly lower than what is due to typical dust.
If their spectral characteristics are not taken into account,
the light curves and color curves of spectroscopically different SNe will be compared like they are coming from a similar object and
 lead to low estimates for the average $R_V$ of SNe~Ia.
An illuminating example of this degeneracy between intrinsic color and dust-extinction 
that can be broken using the spectra is shown by the slow-declining SN\,2007qe \citep[\Deltam$=0.98\pm0.05$,][]{2009ApJ...700..331H}. 
This SN is approximately as slow declining as the bright SN\,1999aa \citep[\Deltam$=0.85\pm0.05$,][]{2006AJ....131..527J}.
 The two light curves are very similar if they are shifted in order to match (Fig.~\ref{fig:Bmag_lc_pred}).
 This degeneracy is not broken using extensive $B-V$ color curve observations.
 Fig.~\ref{fig:BmV_lc_pred} shows how the two $B-V$ color curves are similar. The only difference is a constant shift that is easy to misjudge as reddening.
Looking at the spectra, however, the strong differences between the two SNe are immediately apparent.
The equivalent widths of corresponding lines are significantly different, and the velocity of the lines in the two spectra are extremely different.
SN\,2007qe has a larger velocity and is redder (in agreement with \cite{2011ApJ...742...89F}).
Without the availability of spectroscopic information, the ratio between the color ``excess'' of SN\,2007qe and SN\,1999aa and the difference in luminosity would mimic an extinction with a low $1.<R_V<1.5$.  
This is just an extreme example, and there are many other SNe that are intermediate cases. All these intermediate cases may introduce a shift to the average $R_V$.
\cite{2014ApJ...780...37S} showed that having uncertainty in the determination of the intrinsic color can favour low $R_V$ values.
Using the spectral information, we show that such a dispersion, not accounted by looking only at light curves, is present in the data.

This degeneracy in the light and color curve spaces may be also broken by using properly good $U$ band photometry.
In the $U$ band the effect of dust is the strongest, and from most studies based on color-curve matching it is clear that there is a significant amount of residual intrinsic scatter in this band \citep[e.g][]{2014ApJ...789...32B}.
For example, with pre-maximum U band photometry it may be possible to discriminate between SN\,1999aa and SN\,2007qe.
SN\,1999aa has a bright and early peak in the $U$-band that is not present in SN\,2007qe. However, it may still not be enough to separate intrinsic color from extinction as much as using spectral information for all SNe\,Ia.

The presence of circumstellar dust was proved in a few higly extinguished SNe\,Ia \citep[e.g.][]{2007Sci...317..924P}.
Part of the extinction in some SNe~Ia has to come from circumstellar dust.
If this circumstellar dust is close enough to the SN, multiple scattering produces an extinction with a low $R_V$ \citep{2008ApJ...686L.103G}.
 This can be the source of a low $R_V$ for few highly extinguished SNe.
From a physical point of view, our finding of a normal $R_V$ for most SNe~Ia means that the dust in
front of the majority of them is quite normal.
 This suggests that the
majority of SN\,Ia extinction is caused by simple interstellar dust in the host
galaxy and it is not related to the progenitor of the SN.

\section{Conclusions}
\label{sec:conclusions}

We build on the physically motivated assumption that the intrinsic light curve
and color curve variability are a function of the variability in spectral
series. 

 In the context of SN~Ia, multivariate PLS, together with PCA, becomes
a sharp tool able to separate the intrinsic variability and the variability due
to dust.  This approach can be valuable in a number of challenges, such as the
study of the reddening law and its possible variability and improving the
calibration of SNe~Ia luminosity by the use of their spectra which
have plenty of distance-independent properties.
In this work we investigate the  $R_V$ dust parameter, the most commonly used
parameter to characterize the type of dust. Our results suggest a solution for 
the tension between the range of observed values for this parameter in the
Milky-Way and the values sometimes inferred from SNe\,Ia.
We find that the majority of the host galaxy extinction of SNe\,Ia is caused by normal dust,
similar to the typical Milky-Way dust. This suggests that most of the dust
associated with SNe\,Ia is of interstellar origin and not associated with the
progenitor.

Our technique allows to predict the absolute magnitude of SNe\,Ia without
assuming the decline rate luminosity relation.  This permits to test this
relation without assumptions on the reddening, that depends on how the
intrinsic color of the SN is estimated.  The technique can have a number of
cosmological applications.  Used together with the traditional light curve
fitting, our approach may lead to a better relative calibration of SNe\,Ia with
a good spectral coverage.
 This can be particularly useful to set the relative
calibration between the few SNe\,Ia with a reliable Cepheid distance measurement
and the SNe\,Ia in the smooth Hubble-Flow.

The modelling of SNe\,Ia by means of radiation transport techniques
\citep[e.g.][]{Stehle05} requires precise measurements of the distance and of
the amount of extinction of the selected SN.  Traditional methods have often large
uncertainties on the distance of nearby galaxies and on the host galaxy
reddening. Although the distance of the SN and its intrinsic color can be
obtained directly from the modelling \citep{2014MNRAS.445..711S}, this procedure is time
consuming and introduces new free parameters.  The method described
in our work gives estimates for the absolute magnitude and the intrinsic color
for most common spectral types of SNe\,Ia.  This can be used to speed up their
modelling and reduce the number of free parameters involved. It will also help in keeping 
the consistency of the luminosity and color between the models of distinct SNe.

Our study shows how important is to obtain spectral coverage for the study of SNe\,Ia, not only for the study of peculiar objects, but also for the study of typical SNe\,Ia.
Incoming automated surveys such as Pan-STARRS and the Large Synoptic Survey Telescope (LSST) will produce large amount of light curves without accompanying spectra.
This study suggests investing significant fraction of available telescope time and resourses in studying spectroscopy.
This also suggests that detailed planning of the limited spectroscopic follow-up available could optimize the color and extinction estimation of the entire photometric sample.

Our approach can easily be extended to other colors and
magnitudes in a broader wavelength range. This will be investigated in detail
in future work.

\section*{Acknowledgments}

This work was supported by the Deutsche Forschungsgemeinschaft
via the Transregional Collaborative Research Center TRR 33 ``The
Dark Universe'' and the Excellence Cluster EXC153 ``Origin and Structure
of the Universe''.
EEOI is partially supported by the Brazilian agency CAPES (grant number 9229-13-2).
We thank the anonymous reviewer for the constructive comments, which helped
improve the manuscript.

\label{lastpage}

\appendix

\section[Appendix]{Gaussian Processes}
\label{Appendix}

 The technique assumes that the data are distributed with a
Gaussian distribution with infinite dimensionality. In every epoch the outcome
of a measurement is assumed to follow a Gaussian distribution. Our set of n
photometric measurements can be seen as one realization of an n-variate
Gaussian distribution in an n-dimensional space. Now, two epochs close to each
other are expected to be correlated. That is, the luminosity of our SN does not
change much in a day. Two observations at different epochs are related by a
{\it covariance function}, $k(t_0,t_1)$, that encodes the relation between the
magnitudes at $f(t_0)$ and $f(t_1)$. A safe assumption for the structure of
this covariance is:

\begin{equation}
 k(t_i,t_j)= \sigma_f^2\exp\left[ - \frac{(t_i-t_j)^2}{2\tau^2}\right] +
\sigma_n^2 \delta_{ij} 
\label{eq:k_tt}
\end{equation}

The first term means that the correlation is high for observations distant by
less than $\sim\tau$ and negligible when the time difference is larger, the
second term accounts for the noise. Without noise in the data the
correlation between two realizations temporally very close would be as
high as the covariance of an individual realization.
Adding noise, the covariance of a given measurement is larger by $\sigma_n^2$.
In practice, these coefficients parametrize how slowly the underling function
varies with time $(\tau)$, how big  the total standard deviation of the
magnitudes $(\sigma_f)$ is, and how much uncertainty is due to the noise of the
individual measurements $(\sigma_n)$. $\sigma_f,\sigma_n$, and $\tau$ are
called hyperparameters of the model. For a given set of hyperparameters, it is
possible to produce a large number of realizations that are likely to reproduce
the observations. The average and the standard deviation of these realizations
will look like the fits in Figure \ref{fig:gp_lc}. The quality of the fit,
however, will heavily depend on the choices of the hyperparameters of
equation \ref{eq:k_tt}. For example, if $\sigma_n$ is set to zero the fit will
lie exactly on the photometry overfitting the data. On the other hand, too
large a $\tau$ will remove small scale variations from the fit, flattening the
peaks. The optimal hyperparameters are not chosen by hand but are retrieved
by a maximization of the probability $p(\{\sigma_f,\sigma_n,\tau\}| obs)$ of
having a certain set of hyperparameters for the given observations.

\bibliography{biblio}


\end{document}